\input harvmac
\input epsf

\def\comment#1{}

\Title{\vbox{\baselineskip12pt\hbox{}
\hbox{UGVA-DPT 1998/06-1008}}}
{\vbox{\centerline{Smoothening Transition of Rough}
\vskip2pt\centerline{Surfactant Surfaces}}}
\centerline{M. Cristina Diamantini\footnote{$^*$}{Supported by
an A. v. Humboldt fellowship. On leave of absence from I.N.F.N. and
University of Perugia; e-mail: diamanti@einstein.physik.fu-berlin.de} \ and
\ Hagen Kleinert
\footnote{$^{**}$}{e-mail: kleinert@physik.fu-berlin.de ~~
http://www.physik.fu-berlin/\~{}kleinert}}
\centerline{Institut f\"ur Theoretische Physik, Freie Universit\"at Berlin}
\centerline{Arnimalle 14, D-1000 Berlin 33, Germany}
\bigskip
\centerline{Carlo A. Trugenberger\footnote{$^{***}$}{Supported by a Profil 2
fellowship of the Swiss National Science Foundation;
e-mail:cat@ka\-lymnos.unige.ch}}
\centerline{D\'epartement de Physique Th\'eorique, Universit\'e de Gen\`eve}
\centerline{24, quai E. Ansermet, CH-1211 Gen\`eve 4, Switzerland}

\vskip .3in
\noindent
We propose a model for surfaces in mixtures
of oil, water, and
surfactants with strong electric dipoles.
The dipole interactions
give rise to
a non-local interaction
with a {\it negative stiffness} between surface elements.
We show that, for large space dimension $D$, this model
has a phase transition from  {\it rough\/} to {\it smooth\/} surfaces.
Contrary to models with a simple
quadratic
curvature energy of a positive sign, which always have
a  finite persistence length typical for rough surfaces,
the
smooth surfaces in our model
exhibit
{\it long-range} correlations
of a generalized antiferromagnetic type.
These correlations might be related to the recently
observed ``egg carton" superstructure of membranes.

%\draft
\Date{June 1998}

\lref\gsw{For a review see e.g.:
M. B. Green, J. H. Schwarz and E. Witten, ``{\it Superstring
Theory}", vol. 1, Cambridge University Press, Cambridge (1987).}

\lref\polbook{For a review see e.g.:
A. M. Polyakov, ``{\it Gauge Fields and Strings}", Harwood Academic
Publishers, Chur (1987).}

\lref\polrev{A. M. Polyakov, {\it Physica Scripta} {\bf T15} (1987) 191.}

\lref\pola{A. M. Polyakov, {\it Nucl. Phys.} {\bf B268} (1986) 406.}

\lref\polb{A. M. Polyakov, {\it Nucl. Phys.} {\bf B486} (1997) 23; see also
F. Quevedo and C. A. Trugenberger, {\it Nucl. Phys. } {\bf B501} (1997) 143.}

\lref\ka{H. Kleinert, {\it Phys. Lett.} {\bf B174} (1986) 335
(http://www.physik.fu-berlin/ {}\~{}kleinert/kleinert\_re2\#149).}

\lref\vdW{H. Kleinert, {\it Phys. Lett.} {\bf A163} (1989) 253.}

\lref\sc{H. Kleinert,
{\it Phys. Lett.} {\bf B293} (1992) 168. See also
H. Kleinert, {\it Int. J. Mod. Phys.} {\bf A7} (1992) 4693;
{\it Phys. Lett.} {\bf B246} (1990) 127; Lectures
in Proceedings of a NATO Advanced Study Institute on:
``{\it Formation and Interactions of Topological Defects}",
University of Cambridge, England,
(cond-mat/9503030).}

\lref\kb{H. Kleinert and A. Chervyakov, {\it Phys. Lett. } {\bf B381}
(1996) 286
(http://www.physik.fu-berlin/ {}\~{}kleinert/kleinert\_re2\#241).}

\lref\largedk{H. Kleinert, {\it Phys. Rev. Lett.} {\bf 58} (1987) 1915.}

\lref\larged{H. Kleinert, {\it Phys. Rev. Lett.} {\bf 58} (1987) 1915;
P. Olesen and S. K. Yang, {\it Nucl. Phys.} {\bf B283} (1987) 73;
E. Braaten, R. D. Pisarski and S. M. Tze, {\it Phys. Rev. Lett.}
{\bf 58} (1987).}

\lref\david{F. David and E. Guitter {\it Nucl. Phys.} {\bf B295}
(1988) 332, {\it Europhys. Lett. } {\bf 3} (1987) 1169.}

\lref\bz{E. Braaten and C. K. Zachos, {\it Phys. Rev. } {\bf D35} (1987) 1512.}

\lref\qt{F. Quevedo and C. A. Trugenberger, {\it Nucl. Phys.} {\bf B501}
(1997) 143.}

\lref\dqt{M. C. Diamantini, F. Quevedo and C. A. Trugenberger, {\it Phys.
Lett.}
{\bf B396} (1997) 115.}

\lref\dt{M. C. Diamantini and C. A. Trugenberger,
{\it Phys. Lett.} {\bf B421} (1998) 196;
``{\it Geometric Aspects of
Confining Strings}", hep-th/9803046.}

\lref\dtfut{M. C. Diamantini and C. A. Trugenberger, in preparation}

\lref\kr{V. I.
Ogievetsky and V. I. Polubarinov, {\it Sov. J. Nucl. Phys.} {\bf 4}
(1967) 156; M. Kalb
and P. Ramond, {\it Phys. Rev.} {\bf D9} (1974) 2273.}

\lref\pwz{M. I. Polikarpov, U.-J. Wiese and M. A. Zubkov, {\it Phys. Lett.}
{\bf B309} (1993) 133.}

\lref\orla{K. Lee, {\it Phys. Rev.} {\bf D48} (1993) 2493;
P.Orland, {\it Nucl. Phys.} {\bf B428} (1994) 221; M. Sato and
S. Yahikozawa, {\it Nucl. Phys.} {\bf B436} (1995) 100; E. T. Akhmedov,
M. N. Chernodub, M. I. Polikarpov and M. A. Zubkov, {\it Phys. Rev.}
{\bf D53} (1996) 2087.}

\lref\gr{I. Gradstheyn and I. M. Ryzhik, ``{\it Table of Integrals, Series
and Products}", Academic Press, Boston (1980).}

\lref\wi{E. Witten, {\it Phys. Lett. } {\bf B86} (1979) 283.}

\lref\polcha{For a review see e.g. :
J. Polchinski, ``{\it Strings and QCD}",
contribution in Symposium on Black Holes, Wormholes
Membranes and Superstrings, H.A.R.C., Houston (1992); hep-th/9210045.}

\lref\ps{J. Polchinski and A. Strominger, {\it Phys. Rev. Lett.} {\bf 67}
(1991) 1681.}

\lref\pz{J. Polchinski and Z. Yang, {\it Phys. Rev.} {\bf D46} (1992) 3667.}

\lref\cara{J. C. Cardy and E. Rabinovici, {\it Nucl. Phys.} {\bf B205} (1986)
1.}

\lref\dfj{B. Durhuus, J. Fr\"ohlich and T. Jonsson, {\it Nucl. Phys.}
{\bf B240} (1984) 453; B. Durhuus and T. Jonsson, {\it Phys. Lett.}
{\bf 180B} (1986) 385.}

\lref\hel{
P. B. Canham, {\it J. Theor. Biol.} {\bf 26}, (1970) 61;
W. Helfrich, {\it Z. Naturforsch.} {\bf 28c} (1973) 693;
{\it J. Phys. (Paris)} {\bf 46} (1985) 1263.}

\lref\shape{
H. J. Deuling and W. Helfrich, {\it J. Phys. (Paris)} {\bf 37} (1976) 1335;
W. Harbich,  H. J. Deuling and W. Helfrich,
{\it J. Phys. (Paris)} {\bf 38} (1976) 727.}

\lref\pl{L. Peliti and S. Leibler, {\it Phys. Rev. Lett.} {\bf 54}
(1985) 1690.}

\lref\for{D. F\"orster, ..........}

\lref\kc{H. Kleinert, {\it Phys. Lett. }  {\bf A114} (1986) 263
(http://www.physik.fu-berlin/ {}\~{}kleinert/kleinert\_re2\#128).}

\lref\kd{H. Kleinert, {\it Phys. Lett. } {\bf B211} (1988) 151. See also
M. Kiometzis and H. Kleinert,
      Phys.\ Lett.\ {\bf A140}, 520 (1989), and
the related work in Ref.~\vdW.
}

\lref\dreview{For a review see: F. David, ``Introduction to the Statistical
Mechanics of Random Surfaces and Membranes", in ``{\it Two-Dimensional
Quantum Gravity and Random Surfaces}", D. Gross, T. Piran and S. Weinberg eds.,
World Scientific, Singapore (1992).}

\lref\cpvz{M. N. Chernodub, M. I. Polikarpov, A. I. Veselov and M. A. Zubkov,
``{\it Strings and Aharonov-Bohm Effect in Abelian Higgs Model}",
hep-lat/9804002.}

\lref\trauble{For a review see e.g.:
H. Tr\"auble, ``Membrane Electrostatics", in ``{\it Structure
of Biological Membranes}", S. Abrahamsson and I. Pascher eds., Plenum
Press, New York (1977).}

\lref\kahlweit{M. Kahlweit, R. Strey, R. Schom\"acker and D. Haase,
{\it Langmuir}, {\bf 5} (1989) 305.}

\lref\andelman{D. Andelman, F. Brochard and J.-F. Joanny, {\it J. Chem.
Phys.} {\bf 86} (1987) 3673.}

\lref\egg{M. Antonietti, A. Kaul and A. Th\"unemann,
{\it Langmuir} {\bf 11} (1995) 2633;
B. Kl\"osgen and W. Helfrich, {\it Biophys. Jour.} {\bf 73}
(1997) 3016.}

\lref\spiky{H. Kleinert, FU-Berlin preprint 1998 (cond-mat/9805307) }

Models of random surfaces play an important role in chemical physics,
biophysics, and particle physics. The statistical mechanics of such models
\dreview \ is
usually discussed on the basis of a phenomenological
Hamiltonian which
contains a surface tension
and a positive
extrinsic curvature stiffness
 \hel.
Freely suspended vesicles formed from such
membranes have a vanishing tension,
and their
shapes are determined
by
a positive curvature energy \shape.
Such a curvature energy is known to have
only a finite
persistence length $\xi$
\refs{\pl, \kc, \ka ,\pola }.
The
correlation functions
of the tangent vectors
of the surface fall off
like
$e^{-|x|/\xi}$
signaling a
rough surface
\refs{\pl, \david} \ on  scales larger than $\xi$.
For biomembranes, this limitation
is of no practical importance, since
their persistence lengths
are much larger than the size of the vesicles.
This is not true, however, for layers of surfactants
between oil and water, or for double layers
in water alone.
Their persistence lengths
are of the order of a few hundred \AA.
 Beyond this distance,
membrane
fluctuations are governed
only by tension.
This phenomenon is called {\it spontaneous generation of tension\/}
\refs{\largedk}.
Thus, in these models,
the tension can never disappear completely.
Experimentally, however, it is
quite easy to generate large smooth interfaces.
In fact, this is
done industrially
during the tertiary recovery of oil
from microemulsions. The fractal surfactant interfaces
between oil and water are smoothened
by the addition of salt. This produces a desired
phase separation
in which the oil floats on top of a single interface.
The purpose of this note is to point out a mechanism
which may explain this.

Some years ago it was shown
that the curvature stiffness
of
surfactants at the interfaces between oil and water \kahlweit{}
receives a significant contribution from
the electric fields created by layers and bilayers
of surface
charge \refs{\kd}.
These surfactants
possess electric
dipoles which align along the normals to the interface.
The dipoles can be quite large
\comment{, as e.g. in the case of
the effective dipoles of charged monolayers in the
dipolar regime} \andelman .
The range of the electric fields
can be varied by the addition of salt.
In a field theoretic description,
the dipoles interact
with an electric potential
of a finite mass $\mu $ which is equal to the
inverse
Debye
screening length, thus being
directly governed
by the salt concentration $n$ \trauble :
\eqn\debye{\mu = \sqrt{{8 \pi e^2 n}\over {\epsilon k_B T}}\ ,}
where $k_B$ is the Boltzmann constant,
$e$ the electron charge, and $\epsilon $ the dielectric constant.

In our model we shall ignore the curvature stiffness
generated by the mechanical properties of the
surfactant molecules, or by the van der Waals forces \vdW.
Thus we consider only surfactants with large dipole moments,
where the other forces can be ignored,
and curvature stiffness is {\it completely\/}
due to the dipole-dipole interactions.

The interaction between the dipole moments
in three dimensions is similar to
the tensor interaction between tangent tensors
in  the world surfaces
of models of quark confinement, in which quarks are
held together in the same way as magnetic charges would
in a
superconductor \refs{\sc}.
Similar models have been conjectured  to
describe
the world surface
between quarks
in the confining phase of QCD \polb .
A recent analysis in Refs.~\refs{\dqt, \dt} indicates that
such a model describes smooth surfaces,
a fact which seems to be
confirmed by recent Monte Carlo simulations \cpvz.

By integrating out the electric potential
one obtains a non-local interaction
between normals to the surface.
Unfortunetely, such an interaction is
hard to treat analytically.
For this reason we shall study the phase structure of a simplified version \kb
\ of
this model which retains the essential features of the original one,
most notably a
non-local interaction with a {\it negative stiffness}.

Our starting point is a Hamiltonian describing
surfaces with tension, whose
normal vectors are coupled to
a fluctuating electric field ${E_\mu}(x)=\partial _{\mu } \phi(x)
$:
\eqn\starh{H= r\int d^2\xi \sqrt{g} +
\int d^3 x \ {1\over 2}\ \left( \partial _{\mu} \phi
\partial _{\mu }\phi + \mu ^2 \phi ^2  \right) +
d \int d^2\xi \sqrt{g} \ \partial _{\mu } \phi
\ N_{\mu } \ .}
The surface is
parametrized by the three functions $x_{\mu }(\xi_1, \xi _2)$ ($\mu=1,2,3)$.
The parameter $d$ denotes the
(uniform) dipole density
on the surface, and $\mu$ is the
inverse Debye length \debye.
The intrinsic geometry of the surface
resides in
the
induced metric
\eqn\metric{\eqalign{g_{ab} &= \partial_a x_{\mu } \partial _b x_{\mu }\
,~~~~~~
g = {\rm det} \ g_{ab} \ .\cr }}
The extrinsic geometry is described by the normal vectors
\eqn\normals{\eqalign{N_{\alpha } = {1\over {2\sqrt{g}}}\ \epsilon_{\alpha \mu
\nu }\ &\epsilon _{ab} \ \partial _a x_{\mu } \partial _b x_{\nu }\ ,~~~~~~~
N_{\alpha }N_{\alpha } =1 \ .\cr }}
Here, greek letters refer to the embedding space, while latin letters
denote surface coordinates.
For simplicity we use units in which $\hbar =1$, $c=1$, and
measure all momenta, energies, or distances in units of the microscopic cutoff
$\Lambda $ or its inverse, respectively.
 A natural cutoff is provided by the molecular
size.
We furthermore set
$\Lambda
=1$.

The electric coupling of the surface can be simplified by writing it
in terms of the trace
$C={\rm Tr} \ C_{ab}$ of the second fundamental form, which is
equal to twice the mean curvature:
\eqn\newha{H= r\int d^2\xi \sqrt{g}
+\int d^3 x \ {1\over 2}\ \left( \partial _{\mu} \phi
\partial _{\mu }\phi + \mu ^2 \phi  ^2  \right) +
d \int d^2\xi \sqrt{g} \ \phi C \ .}
Integrating out the electric field, we obtain
the non-local energy
\eqn\nonlo{\eqalign{H= r\int d^2\xi \sqrt{g}
-{d^2\over 2} \int d^2\xi
\int d^2\xi ' \ &\sqrt{g} C(\xi )
\ Y\left( x(\xi)-x(\xi ') \right)
\ \sqrt{g'} C(\xi ') \ , }}
where
\eqn\nonlon{
Y(x) ={{\rm e}^{-\mu |x|} \over 4\pi |x|}\ . }
is the Yukawa potential.

We then proceed by rewriting the non-local
kernel in the embedding space as a corresponding
non-local kernel on the surface. To this end we introduce a new local
coordinate
system around each point of the surface: $\xi^1, \xi^2, \chi^3$.
The coordinates $\xi^1$ and $\xi^2$
at
 $\chi ^3=0$
describe the original surface. Together with the coordinate $\chi ^3$,
they form a
locally flat coordinate system orthogonal to the surface. The coordinate
transformation is described by functions
$x_{\mu }(\xi^1, \xi^2, \chi^3 )$
which for
 $\chi ^3=0$ coincide with the original
parametrization $x_{\mu }(\xi^1, \xi^2)$
of the
surface.

The Yukawa potential in \nonlo \ can be rewritten as
\eqn\yua{\int d\chi
\int d\chi ' \ \delta (\chi) \ {1\over {\mu ^2-\nabla^2}} \ \delta^3
\left( x (\xi , \chi)- x (\xi ' ,\chi ')\right)
\ \delta (\chi ') \ .}
We can now use the transformation rules
\eqn\trru{\eqalign{\delta^3 ( x- x') &= {1\over \sqrt{g}}\ \delta^2
(\xi -\xi ') \ \delta (\chi - \chi ') \ ,\cr
\mu ^2-\nabla ^2 &= M^2- \nabla^2_{\chi}\ ,\cr
M^2 &\equiv \mu ^2-{\cal D}^2\ ,\cr
{\cal D}^2 &= {\cal D}^a{\cal D}_a= {1\over \sqrt{g}}
\ \partial_a g^{ab} \sqrt{g} \ \partial_b \ ,\cr }}
where ${\cal D}_a$ denotes covariant derivatives along the surface, to rewrite
the
Yukawa potential as
\eqn\yub{\eqalign{Y\left( x(\xi )-
x (\xi ')\right)
&= \int d\chi \ \ \delta(\chi) {1\over {M^2-\nabla^2_{\chi}}}
\delta (\chi ) \ {1\over \sqrt{g}}\ \delta^2(\xi - \xi ') \cr
&= G \left(
 {\cal D}^2 \right)
\ {1\over
\sqrt{g}}\ \delta^2(\xi -\xi ') \ ,\cr}}
where $G$ is formally defined by an
expansion in powers of $({\cal D}/\mu )^2$.
In order to compute this,
we first note that
$(M^2-\nabla^2_{\chi})^{-1} \delta (\chi )$
is the Yukawa Green's function in one dimension, and thus equal to
 ${\rm exp}\left( - M|r| \right)/2M $.
The first delta function in
\yub \ tells us then that we have to take this function at $r=0$,
implying that
\eqn\vdt{G \left({\cal D}^2\right)
={1\over 2\mu } \ {1\over \sqrt{1 -
\left( {\cal D}/ \mu \right) ^2 }} \ .}
Inserting this result into \nonlo , we obtain
\eqn\hos{\eqalign{H &= r\int d^2\xi \sqrt{g}
-{d^2\over 4\mu } \int d^2\xi \sqrt{g} \ C
{1\over \sqrt{1-\left( {{\cal D}/ \mu } \right) ^2}} C \cr
&= r
\int d^2 \xi \sqrt{g} -{d^2\over 4\mu } \int d^2\xi \sqrt{g}
\ C^2 + \dots \ .\cr }}
This shows that dipole dominance leads to a non-local interaction
with {\it negative} stiffness
\eqn\stiff{\kappa ={d^2\over 4\mu }\ .}

The formulation \hos \ is rather awkward for analytic computations due
to the non-linear character of the trace $C$.
For this reason
we shall investigate a
model \kb , which
is simpler to handle while
embodying the most
important features of \hos , namely non-locality and negative stiffness.
This
model is formulated in terms of the tangent vectors ${\cal D}_a x_{\mu}$
and posseses the following Hamiltonian:
\eqn\newa{\beta H = \int d^2{\xi } \sqrt{g}\ \ g^{ab}
{\cal D}_a x_{\mu } \ W\left( z , m ,
{\cal D} ^2
\right) \ {\cal D}_b x_{\mu } \ ,}
Here the index $\mu $
runs over $\mu =1,\dots ,D$, with $D$ the dimension of the embedding space,
which we keep variable from now on. The interaction kernel
is
\eqn\intera{\eqalign{W(z , m, {\cal D}^2) &=
{z \over {1-{{\cal D}^2/ m^2}}} \ ,~~~{\rm where}~~
z = {\beta r\over 2}\ ,~~~
m = {\sqrt{2r\mu }\over d}\ .\cr }}
It has the expansion
\eqn\extra{W \left( z , m,
{\cal D}^2 \right)
= z + s {\cal D}^2 + \dots \ ,}
with $s=z/m^2= \beta \kappa $ being the reduced dimensionless stiffness
parameter.
These expansion terms correspond to the
Hamiltonian
\eqn\expaha{H=r\int d^2\xi \sqrt{g} + \kappa
\int d^2{\xi } \sqrt{g}\ \ g^{ab}
{\cal D}_a x_{\mu } {\cal D}^2
{\cal D}_b x_{\mu } + \dots \ ,}
which matches
(up to boundary terms)
\hos \ to this order.

This model can now be analyzed
non-perturbatively by
standard large-$D$ techniques along the lines of
Refs.~\refs{\largedk,\larged,\david} .
To this end we introduce a dimensionless
Lagrange multiplier matrix $\lambda ^{ab}$
to enforce the constraint $g_{ab}=\partial _ax_{\mu }\partial_bx_{\mu }$,
\eqn\lamu{\beta H \to \beta H + \int d^2\xi \sqrt{g} \ \ \lambda
^{ab} \left( \partial _a x_{\mu } \partial _bx_{\mu } - g_{ab} \right) \ .}
We then parametrize the world-sheet in the Gauss map as
\eqn\gauss{x_{\mu } (\xi ) = \left( \xi _1, \xi _2, \phi ^i (\xi )
\right) \ ,\qquad \qquad i=3, \dots , D\ ,}
where $-R_1 /2\le \xi_1 \le R_1 /2$,
$-R_2 /2 \le \xi ^2 \le R_2 /2$ and
$\phi ^i(\xi )$ describe the $D-2$
transverse fluctuations.
Then we search for an isotropic saddle point
of the form
\eqn\iso{g_{ab}=\rho \ \delta_{ab} \ ,\qquad \qquad \lambda ^{ab} =
\lambda \ g^{ab} \ ,}
for the metric and the Lagrange multiplier of infinite systems
($R_1 ,R_2 \to \infty $). At such a saddle
point, we obtain a Hamiltonian
\eqn\gmapac{\beta H =2\int d^2\xi \ \left[ z +\lambda
(1-\rho ) \right] + \int d^2\xi  \ \partial_a\phi ^i
\left[ \lambda + W\left( z, m , {\cal D}^2
\right) \right] \ \partial_a \phi ^i\ .}
Integrating out the transverse fluctuations, we obtain in the infinite-area
limit the free energy
\eqn\newact{\beta F = 2A_{\rm ext} \ \left( z +\lambda
(1-\rho ) \right) + A_{\rm ext} {{D-2}\over 8\pi^2 }\rho
\int d^2p\ {\rm ln}
\left\{ p^2 \left[ \lambda + W
\left( z, m, p^2\right) \right] \right\} \ ,}
where $A_{\rm ext}=R_1 R_2$ is the extrinsic, physical area. For large
$D$, the fluctuations of $\lambda $ and $\rho $ are suppressed and these
variables take their classical values, determined by the two saddle-point
equations
\eqn\sapoi{\eqalign{\lambda &= {{D-2}\over {8\pi }}\ \int_0^1dp\ p\ {\rm ln}
\left\{ p^2 \left[ \lambda + W
\left( z, m, p^2\right) \right] \right\} \ ,\cr
{{\rho-1}\over \rho} &= {{D-2}\over{8 \pi }}
\ \int _0^1 dp\ p\
{1\over {\lambda + W \left( z, m, p^2\right) }}\ ,\cr }}
where we have introduced the ultraviolet cutoff
by restricting the momentum integrations to $p<1$.
Inserting the first saddle-point equation into \newact,
we find
\eqn\efac{\beta F = 2\left( z+\lambda \right)
\ A_{\rm ext} \ ,}
showing that
the tension is renormalized to
\eqn\tension{\alpha = 2(z+\lambda )\ .}

The physics of our surfaces in the large-$D$ limit is determined thus
by the two saddle-point equations \sapoi . The first of these equations
requires the vanishing of the {\it saddle-function\/}
\eqn\safu{f(z, m, \lambda) \equiv \lambda - {{D-2}\over {8\pi }}
\ \int_0^1dp\ p\ {\rm ln}
\left\{ p^2 \left( \lambda + W
\left[ z, m, p^2\right) \right] \right\} \ ,}
and determines the Lagrange multiplier $\lambda $ as the solution of a
transcendental equation. After this, the second saddle-point equation
determines the metric via the equation
\eqn\lame{\rho = {1\over f'(z, m, \lambda )}\ ,}
where a prime denotes the derivative with respect to $\lambda $.

The simple interaction \intera \ in \newa \ has
the advantage that the saddle function
\safu \ can be computed exactly:
\eqn\gapkle{\eqalign{f(z, m, \lambda ) =
\lambda &- {{D-2}\over 16 \pi}\left\{ -1
-m^2 \ {\rm ln}\left( 1+{1\over m^2}\right) - {{m^2 (\lambda +z)}
\over \lambda } \ {\rm ln}\left( {{m^2 (\lambda +z) }\over {1+m^2}}
\right) \right\} \cr
&- {{D-2}\over 16 \pi}\left\{
{{m^2 (\lambda +z) +\lambda }\over \lambda }
\ {\rm ln} \left( \lambda +{z m^2 \over {1+m^2}} \right) \right\} \ .\cr }}
This function has the following limiting values:
\eqn\prope{\eqalign{\lim _{\lambda \to \infty} f(z, m, \lambda )
& =\infty \ ,\cr
\lim _{\lambda \to \lambda _{\rm min}} &= -z m^2 -{{D-2}\over {16 \pi }}
\left(
-1+{\rm ln} \ z m^2 \right) + {\cal O}\left( m^2 \ {\rm ln} \ m^2 \right) \ ,
\qquad m\ll 1\ ,\cr
\lambda _{\rm min} &= {-z m^2 \over {1+m^2}} \ .\cr }}
For $z$ sufficiently large we have $\lim _{\lambda \to \lambda _{\rm min}},
<0$ and there exists at least one solution to the saddle-point equation
$f(z, e, \lambda )=0$.

The derivative of the saddle-function $f$,
\eqn\dsf{f'(z, e, \lambda ) = 1-{{D-2}\over {16 \pi }}\left\{ {1\over \lambda}
-{zm^2 \over \lambda ^2} \left[ {\rm ln}\left( 1+{1\over m^2} \right) +
{\rm ln} {{\lambda +{zm^2 \over {1+m^2}}} \over
{\lambda +{zm^2 \over m^2}}}
\right] \right\} \ ,}
determines the metric element $\rho $ via \lame . Given that
\eqn\proder{\eqalign{\lim _{\lambda \to \infty} f'(z, m,\lambda ) &= 1\ ,\cr
\lim_{\lambda \to \lambda_{\rm min}} f'(z, m, \lambda) &= -\infty \ ,\cr }}
the saddle-function $f$ must have an odd number of extrema.
Our numerical analysis shows that it has
exactly one minimum. When this minimum lies
above zero, the saddle-point equations have no solutions. When the minimum
lies below zero the saddle-point equation $f(z, m, \lambda)=0$ has two
solutions. Only the largest of these two solutions, however, is physical since
at the smallest one $f'(z, m, \lambda )=1/\rho <0$. Then
we have exactly one physical solution of the saddle-point equations.

In Fig.1
we plot the critical line in parameter space below which there exists
one solution to the saddle-point equations for $D=3$. We choose to plot this
line as a function of the parameters $m$ and $k_B T/r = 1/2z$, where  $T$
is the temperature.

\vskip 30 pt

\epsfbox{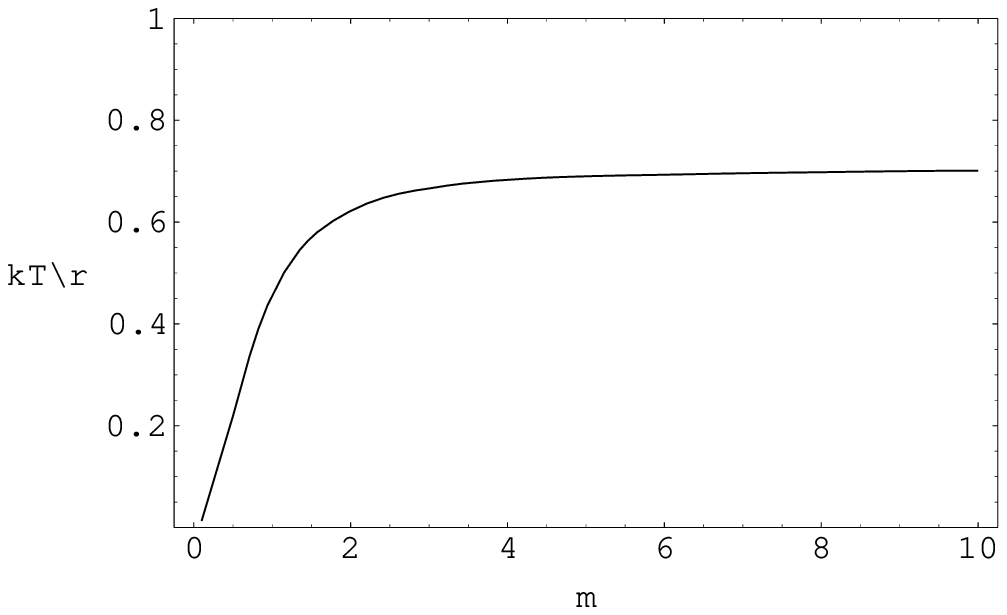}

\vskip 3pt
{\centerline {Fig. 1: Critical line for
roughening
transition
at
large-$D$.}}
\vskip 30pt

\noindent
When the critical line is approached from below,
the minimum of the saddle-function
$f(z, m, \lambda )$ goes to zero, so that $\lambda \to \lambda ^*$
which is the solution of the equation  $f' \left( z, m, \lambda ^* \right) =0$.
Using \lame, we
conclude that
\eqn\phtr{\lim_{(z, m) \to (z, m)_{\rm cr}} \rho \ \ = \infty \ .}
Thus, when approaching to the critical line, the {\it intrinsic area}
of the surface {\it diverges}
whereas the renormalized tension remains finite. As far as the phase transition
with respect to $T$ or $1/r$ is concerned,
 this property is
also found
in the
model
with positive stiffness \david . In the
low-temperature phase of that
model, however, there are short-range correlations
between normals to the surface, indicating roughness
\refs{\pl, \david}.
As we show below,
the situation is totally different in our model, as a consequence of
the negative stiffness. Note that our model reproduces correctly
the positive slope of the critical line in ($T$, $n$)-space, typical
of ionic surfactants \kahlweit .

The geometric aspects of the surface are embodied in the correlation
function
\eqn\goas{g_{ab}(\xi -\xi ') = \langle \partial_a \phi ^i (\xi )
\ \partial_b \phi ^i (\xi ') \rangle \ ,}
for the normal components of the tangent vectors.
This is immediately obtained from \gmapac \ as
\eqn\ladcorr{g_{ab}(\xi -\xi ')
=\delta _{ab}
\ {1\over (2\pi )^2}\ \int d^2p \ {1\over
2\left[ \lambda + W\left( z, m, p^2\right) \right] }
\ \ {\rm e}^{i \sqrt{\rho } p (\xi -\xi ')} \ ,}
where $W\left( z, m, p^2\right) $
is the Fourier transform of the
interaction \intera . This correlation function can
be rewritten as
\eqn\newcorr{g_{ab}(\xi -\xi ')
=\delta _{ab}
\ {1\over (2\pi )^2}\ \int d^2p \ \left\{
{1\over 2\lambda} + {z\over {|\lambda | \left( \alpha
-2s{|\lambda |\over z} p^2 \right) }} \right\} 
\ \ {\rm e}^{i \sqrt{\rho } p (\xi -\xi ')} \ ,}
where we have used the fact that the solution for the
saddle-point value of $\lambda $ is negative. 
The dominant large-distance behaviour of
this correlation function is given by:
\eqn\larlar{g_{ab}(\xi -\xi ') \simeq \delta_{ab}\ {z^2\over {8s\lambda ^2}}
\ \sqrt{2\over {\pi \sqrt{{z \alpha \rho} \over 2s |\lambda |} |\xi -\xi '|}}
\ {\rm sin} \left( \sqrt{{{z \alpha \rho} \over 2s |\lambda |}}
|\xi -\xi '| - {\pi\over 4}\right) \ .}
It is easiest to check this result
by computing backwards its Fourier transform.
To this end it is useful to recognize
that \larlar \ represents the asymptotic behaviour of the
von Neumann function
$(z/8s \lambda ^2) N_0\left( \sqrt{z \alpha \rho /2s|\lambda |} |\xi -\xi '|
\right) $. Computing the two-dimensional
Fourier transform of this function one finds
\gr \ $z/|\lambda |(\alpha -2s (|\lambda |/z) 
p^2)$ for all $0<p<\sqrt{z\alpha /2s|\lambda |}$
i.e. one reproduces exactly
the correlation \ladcorr \ in momentum space.

The correlation function \larlar \ is valid up to a large
infrared cutoff $R$ such that (1/R) provides a regulator
for the pole at $p=\sqrt{z\alpha /2s|\lambda |})$ in \newcorr .
So, we have {\it long-range} order in the sense that there
is no finite correlation length and  
$\int d^2\sigma \ g_{aa}(\xi ) $ diverges like $\sqrt{R}$. The
oscillatory behaviour is due to the negative stiffness: it indicates
that the surface fluctuates around the 
reference plane with an intrinsic wavelength
\eqn\wale{\ell \propto \sqrt{s |\lambda |\over {z \alpha}} \ .}
These long-range correlations are thus of a generalized antiferromagnetic type,
with domains of orientational order of size $\ell $. This is a
consequence of the frustrated antiferromagnetic interaction
with negative stiffness \dt .

Contrary to models with a positive stiffness, our model predicts the existence
of a smooth surface phase with long-range correlations for small $T/r$ and
large $m$.
Notice that, due to \phtr \ the correlations
$\langle \partial_a \phi ^i (\xi ) \ \partial_b \phi ^i (\xi ') \rangle $
vanish on the critical line.te

We don't know the exact nature of the
{\it roughening transition} occuring on
the critical line depicted in fig. 1. The diverging intrinsic area and
the vanishing correlators, however, indicate
for certain that the rough surface
is a fractal object.
Note, however, that it is not of the type expected
for a
crumpled surface, where we
should have a homogeneous solution to the saddle-point equations,
while
the correlation functions \ladcorr \ would be short-range. It is not clear
 to us whether
a generic solution of the saddle-point equations, not restricted by the
ansatz \iso , exists beyond the critical line. Such a solution would
correspond to a fractal phase where the surface develops fingers and bumps.
Also, it is not clear whether the large-$D$ transition survives the
corrections to $D$=3. If not, the surfaces would always be in the smooth,
perturbative phase. In any case, we have shown how dipole dominance
provides a viable mechanism for large, smooth surfactant interfaces.
In addition, we have shown how a non-local interaction with negative
stiffness leads generically to long-range, periodic correlation functions.
Thus, our model might be of relevance for
the recently observed \egg \ ``egg carton"
superstructure of membranes,
although there are also other possible mechanisms
to explain these \spiky.

\listrefs
\end